\documentclass[aps,prl,showpacs,twocolumn]{revtex4}
\usepackage{graphicx}
\usepackage{amsmath}
\begin{document}
\title{Attosecond spectroscopy of solids: streaking phase shift due to lattice scattering}  
\author{E. E. Krasovskii}
\affiliation{%
Department of Materials Physics, University of the Basque Country UPV/EHU,
20080 Donostia-San Sebasti\'an, Spain
}
\affiliation{%
IKERBASQUE, Basque Foundation for Science, 48011 Bilbao, Spain
}
\affiliation{%
Donostia International Physics Center (DIPC), 
20018 Donostia-San Sebasti\'an, Spain
}

\begin{abstract}
Theory of laser-assisted photoemission from solids is developed 
for a numerically exactly solvable model with full inclusion of 
band structure effects. The strong lattice scattering in the vicinity 
of band gaps leads to a distortion and a temporal shift of the streaking 
spectrogram of the order of 100~as. The effect is explained in terms of 
Bloch electron dynamics and is shown to remain large for an arbitrarily 
small photoelectron mean free path. The implications for the streaking 
experiment on W(110) 
[A.~L.~Cavalieri {\it et al.}, Nature (London) {\bf 449}, 1029 (2007)]
are discussed.                                               
\end{abstract} 
\maketitle 

Attosecond photoelectron streaking is a unique tool to study ultrafast 
electronic processes with subfemtosecond resolution~\cite{KI09}. The
method relies on mapping time to energy by means of a strong laser field: 
the electron wave packet created by an ultrashort pulse of extreme 
ultraviolet radiation (XUV) is accelerated by the superimposed laser 
field, and its energy spectrum $J(\epsilon)$ shifts up or down depending 
on the momentum transfer from the laser field to the outgoing electron. 
The energy gain depends on the electron release time $\tau$ relative to 
the laser field $E_{\textsc l}(t)$, as encoded in the spectrogram 
$J(\epsilon,\tau)$. In comparison to the isolated atom, the application 
of the method to solids is complicated by two factors. First, the outgoing 
wave is a coherent superposition of contributions from many atoms. Second, 
the crystal lattice strongly affects the propagation of the photoelectron.

In the seminal experiment on W(110) by Cavalieri {\it et al.}~\cite{Ca07} a 
shift of $\Delta\tau=110\pm 70$~as was discovered between the spectrograms of 
valence 5$d$ and semi-core 4$f$ states. Its origin has been debated in recent 
theoretical studies~\cite{BM08,Le09,KE09,ZT09}. In Refs.~\cite{BM08,Le09,KE09}
the original idea~\cite{Ca07} is developed that the shift is due to a spatial 
inhomogeneity of the streaking field: under the assumption that the laser 
field does not penetrate into the crystal the measured $\Delta\tau$ is just 
the difference in the travel time to the surface for 5$d$ and 4$f$ electrons. 
Baggesen and Madsen~\cite{BM08} proposed to use the states located outside 
the metal as a reference to measure the absolute travel times. In the classical 
simulation by Lemell {\it et al.}~\cite{Le09} the streaking phase shift was 
traced back to the escape depth and to the electron group velocity. Kazansky 
and Echenique~\cite{KE09}, however, questioned the applicability of the concept 
of group velocity as derived from band structure and ascribed the delay both to 
the different character of the initial 5$d$ and 4$f$ states and to a different 
propagation of the outgoing electrons. Contrary to the other authors, Zhang and 
Thumm~\cite{ZT09} assumed the laser field to be spatially constant up to the 
escape depth of 5~{\AA}, but, similar to Ref.~\cite{KE09}, found the difference 
in the localization of the initial states to be the reason for their different 
streaking behavior. One important aspect not considered in previous studies 
is the Bloch nature of the outgoing electron and its implications for the 
streaking spectrogram.

Knowledge of the crystal lattice scattering effects is essential for the 
understanding of what is being measured in a streaking experiment on a 
solid. The aim of the present Letter is to shed light on how the band 
structure affects the formation of the spectrogram. Its role is twofold: 
First, close to the band gaps, the group velocity of the outgoing electron 
is reduced, which reduces the energy transfer to the photoelectron even if 
the laser field in the solid is as strong as in vacuum. Second, if the final 
state is close to a band gap edge, an electron at the surface is brought by 
the XUV pump pulse towards the crystal interior before it goes back and 
leaves the solid. This complicated trajectory causes a distortion of the 
spectrogram, which can be perceived as a final-state-energy dependent delay 
in photoemission. The joint effect of the two factors is shown to result in 
values of $\Delta\tau$ of the order of 100~as, i.e., close to those 
experimentally observed.

\begin{figure}[t]   
\begin{minipage}{0.48\textwidth}
\includegraphics[width=0.385\textwidth]{fig1a}
\hfill 
\includegraphics[width=0.570\textwidth]{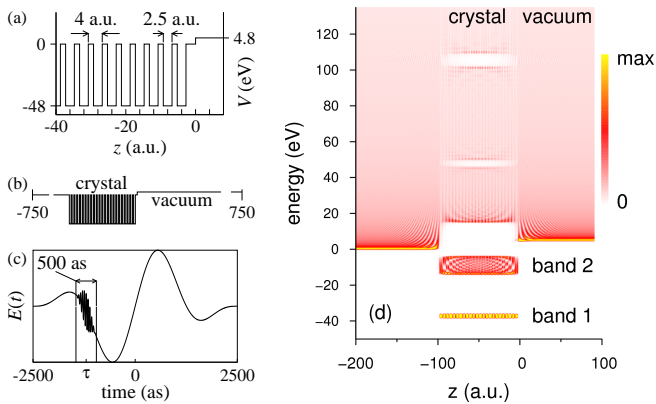}
\end{minipage}
\caption{\label{box} (color online) 
(a) Potential profile of the model crystal. Atomic layers are given 
by potential wells, lattice constant is 4~a.u. (b)
Domain of $\hat H$:  $-750 \le z \le 750$~a.u. (c) Sum of XUV and laser 
pulses. (d) $z$-resolved density of states (DOS). Energy 
is relative to the ``muffin-tin'' (MT) zero. 
}
\end{figure}

To arrive at reliable conclusions on the dependence of $\Delta\tau$ 
on the final state energy it is necessary to avoid any uncontrollable 
numerical approximations. We consider a model crystal surface without 
lateral corrugation, for which the time-dependent Schr\"odinger equation 
(TDSE) can be solved and the spectrogram $J(\epsilon,\tau)$ can be obtained
with any desired accuracy. The periodic lattice is represented by a slab 
of 24 atomic layers with a piecewise constant potential $V(z)$, see 
Fig.~\ref{box}(a). We consider normal emission and assume a linear 
polarization of the light with electric field along the surface normal, 
so the problem reduces to a one-dimensional (1D) TDSE. The perturbing 
field $E(t)$ is a superposition of an XUV and a laser pulse, see 
Fig.~\ref{box}(c). The temporal envelopes of both pulses are of the form 
$\cos^2(\pi t/D)$. The full duration of the XUV pulse is $D=500$~as. The 
laser pulse is chosen to be an odd function of $t$: 
$E_{\textsc l}(t)=E_{\textsc l}^{\textsc m}\sin(\Omega t)\cos^2(\pi t/D)$,
with photon energy $\Omega=1.65$~eV, $D=5$~fs, and amplitude 
$E_{\textsc l}^{\textsc m}=2\times 10^7$~V/cm. The system is enclosed in a 
box [Fig.~\ref{box}(b)], and all calculations are performed in 
matrix form in terms of exact eigenfunctions (discrete spectrum) of the
unperturbed Hamiltonian $\hat H=\hat p^2/2m+V(z)$. The electric field 
perturbation is taken in the dipole approximation, $\Delta\hat H =zeE(t)$. 
Note that the XUV pulse is not separated from the laser pulse -- input 
into $\Delta\hat H$ is their sum. The TDSE is solved with the split-operator 
technique in the spectral representation, so the effect of the crystal 
potential is fully taken into account for both initial and final states. 
The spectrum of $\hat H$ is truncated at 200~eV above the vacuum level, 
which ensures the convergence of all the results. More details of the 
methodology can be found in Ref.~\cite{KB07}, where the model was applied 
to a single atom. 

Initial states are the two bound bands, see Fig.~\ref{box}(d): the
narrow band (1) and the wide band (2), respectively, model semi-core 
and valence states. Each initial state gives rise to two wave packets
traveling in opposite directions. When both pulses are over and the
right-traveling packet has left the crystal, it is reexpanded in terms
of the eigenfunctions of $\hat H$ to yield the recorded spectrum.
The observed spectrum $J(\epsilon,\tau)$ is a sum of the spectra of
the 24 levels representing the band. The centroid (center of gravity)
of the total streaked spectrum $\tilde\epsilon(\tau)$ shifts relative
to the centroid of the laser-free spectrum $\tilde\epsilon_0$, and the
dependence of the shift on the release time $\tau$ is referred to as
the streaking curve, $s(\tau)=\tilde\epsilon(\tau)-\tilde\epsilon_0$. 
Note that for a classical particle created at the time $\tau$ and moving 
under the field $E_{\textsc l}(t)$ in a flat potential $V(z)=\text{const}$, 
the streaking curve is symmetric: $s(\tau)=s(-\tau)$.

\begin{figure}[b]    
\begin{minipage}{0.4\textwidth}
\includegraphics[width=\textwidth]{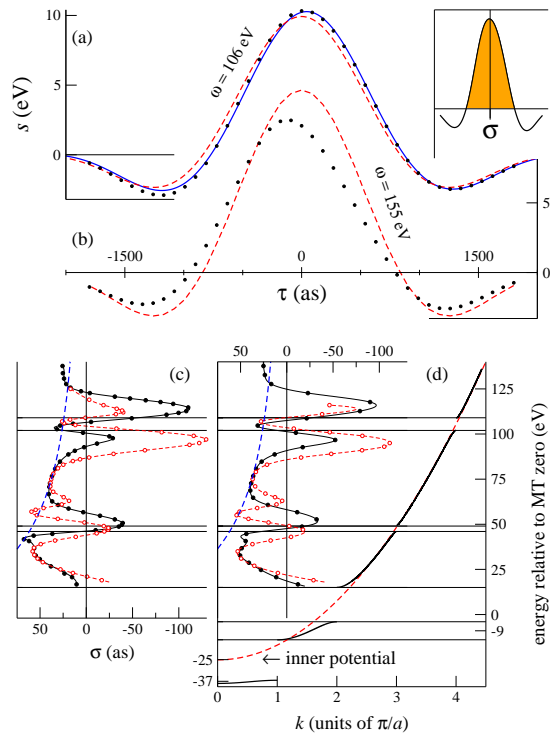}
\end{minipage}
\caption{\label{band_streaking} (color online) 
(a) Blue solid line is for a classical particle traversing the potential barrier
$U$. 
(a) and (b) Circles are the $s(\tau)$ values by TDSE for the emission from band 1 
($\epsilon_{\rm ini}=-37$~eV) with $\omega=106$~eV (a) and 155~eV (b). Red dashed 
lines are for a classical particle (starting with the same kinetic energy) in a 
flat potential. Inset illustrates the definition of $\sigma$.
(c) and (d) Streaking curve shift $\sigma(\epsilon)$ for a classical particle 
(thick dashed line) and for emission from band 1 (black full circles) and band 2 
($\epsilon_{\rm ini}=-9\pm5$~eV, red open circles). 
$\epsilon=\epsilon_{\rm ini}+\omega$. Laser field is 
$E_{\textsc l}^{\textsc m}= 2\times 10^7$~V/cm (c) and $4\times 10^7$~V/cm (d). 
Shown to the right are the
crystal band structure $\epsilon(k)$ and the parabolic fit
to the unbound spectrum (red dashed line) to determine the inner potential $U$.
}
\end{figure}

Let us start with the obvious observation that the average potential in the 
crystal $U$ is lower than the potential in vacuum, so in the crystal a classical 
electron moves faster than in vacuum. Thereby its streaking curve deviates from 
that of a free particle, and this distortion can be perceived as a delay of the 
release of the photoelectron coming from the depth of the crystal relative to the 
one released just at the surface. For example, in the present case, the momentum 
transfer for a free electron would be maximal at $\tau=0$, but because at that 
point the electric field increases the electron in the crystal gains more energy 
if it starts later to be exposed to a stronger field while it moves faster. 
Figure~\ref{band_streaking}(a) shows the exact result for a classical particle 
(full line) starting from the depth of 55~a.u. with the initial kinetic energy 
of 94~eV in a solid with an inner potential of $U=-30$~eV. This corresponds to 
the excitation of the narrow band by the XUV pulse of $\omega=106$~eV 
($\epsilon=69$~eV), and the fully quantum-mechanical calculation yields an 
almost identical streaking curve: it is seen to be shifted to the right by 
$\sim 50$~as from the free-particle curve~\cite{Na11}. What happens when the 
final state is close to a band gap, where the electron group velocity in the 
crystal is lower than in vacuum? Here, by the Bloch electron dynamics law 
$\hbar dk/dt=F$, its energy changes slower than in vacuum, and the opposite 
effect is expected. Indeed, for $\omega=155$~eV ($\epsilon=118$~eV) the 
$s(\tau)$ curve is shifted to the left, see Fig.~\ref{band_streaking}(b), 
i.e., the electron appears to be released earlier than the free particle.

For the special case of an odd function $E_{\textsc l}(t)$ the temporal shift 
of the curve can be roughly quantified as the center of gravity $\sigma$ of 
the positive part of the function $s(\tau)$. Its dependence on the final 
energy $\sigma(\epsilon)$ for the classical particle and for both bands is 
shown in Figs.~\ref{band_streaking}(c) and~\ref{band_streaking}(d). Far from 
the band gaps the streaking curve shows a delay, and $\sigma$ is close to that 
for a classical particle starting at $-55$~a.u., which is slightly deeper than 
the average spatial location of the initial states, $-48$~a.u. When the final 
state energy approaches a band gap the $s(\tau)$
curve strongly deviates from the free-electron shape, and the function 
$\sigma(\epsilon)$ shows oscillations. This is caused by rapid energy 
variations of the photoemission cross-section in combination with the large 
energy width of the pump pulse~\cite{cmt1}, so that the spectrum is composed 
of several contributions, which are differently streaked by the probe field. 
Nevertheless, when the spectral weight concentrates close to a band-gap edge, 
the $s(\tau)$ curve retains an approximately symmetric shape and shifts to the 
left, showing an advance (negative $\sigma$). The $\sigma(\epsilon)$ curves 
are rather stable to the strength of the streaking field: a two times larger 
amplitude $E_{\textsc l}^{\textsc m}$ yields almost the same result 
[cf. Figs.~\ref{band_streaking}(c) and \ref{band_streaking}(d)]. This proves 
the robustness of the effect and that the parameter $\sigma$ is a reasonable
choice to characterize its range. The $\sigma(\epsilon)$ curves are very 
similar for the emission from the narrow and from the wide band, with more 
oscillations in the latter case due to the much broader spectrum of the wide 
band. Far from the gaps the values of $\sigma$ for the two bands closely 
agree, demonstrating the dominant role of final states.

We may conclude that the semi-classical picture based on the notions of inner 
potential and group velocity is a good starting point to understand the
$\tau$-shift of the streaking curve. To relate it to the experiment, we 
must take into account the surface sensitivity of photoemission: owing 
to inelastic scattering, only the electrons excited in a close vicinity 
of the surface contribute to the observed spectrum. The travel path to 
leave the crystal is, thus, limited by the escape depth $\lambda$, and 
for sufficiently small $\lambda$ the effect of {\em delay} that occurs 
in nearly-free-electron (NFE) regions (far from band gaps) must become 
negligible. However, the effect of {\em advance} would 
not completely disappear even for a small $\lambda$ of few Angstroms because 
close to the gap the velocity may be arbitrarily small. For an unambiguous 
analysis of the escape depth effect it is desirable to avoid a phenomenological
treatment of inelastic effects by an imaginary absorbing potential
and current non-conservation. To introduce a spatial resolution into the 
microscopic theory, let us create a defect in the 1D lattice and study the 
streaking of the photoelectron from the localized state at the defect.

\begin{figure}[t]    
\begin{minipage}{0.48\textwidth}
\includegraphics[width=\textwidth]{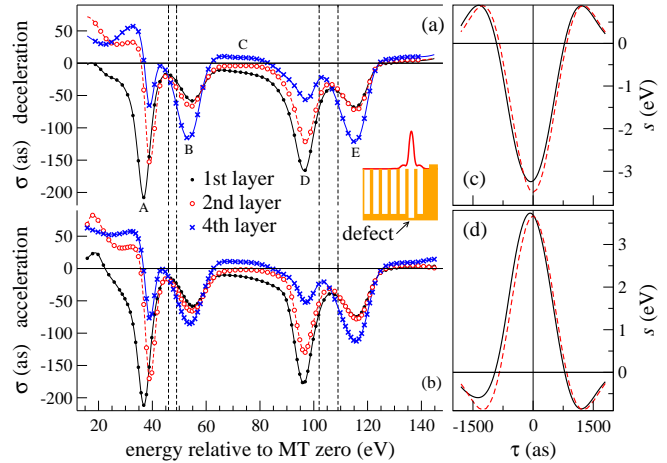}
\end{minipage}
\caption{\label{defect} (color online) (a) and (b) Delay $\sigma(\epsilon)$ for 
emission from the state at the defect (inset) in the 1st (black full circles), 
2nd (red open circles), and 4th layer (blue crosses). 
$\epsilon_{\rm ini}=-41$~eV, $\epsilon=\epsilon_{\rm ini}+\omega$. 
(c) and (d) $s(\tau)$ for $\omega=82$~eV for the defect in the 1st layer 
(full lines) and for a classical particle in a flat potential (dashed lines).
 Vertical lines show band gaps.
(a) and (c) are for the laser pulse $E_{\textsc l}(t)$; (b) and (d) for
$-E_{\textsc l}(t)$.
}
\end{figure}

Figure~\ref{defect} shows the energy dependent distortion of the streaking 
curves for the initial state localized at the defect in the first, second, 
and fourth layer. In accord with the above discussion, the effect of delay 
in the NFE regions is rather small (although it increases with the depth). 
However, the effect of advance is quite large: as in the case of the
extended states, the variations of $\sigma$ are comparable to the measured 
$\Delta\tau$ value. The most surprising is that for the 1st layer $\sigma$ 
does not vanish -- it is of the same order or even larger than for the 2nd 
and the 4th layer. Although for the electron released at the 1st layer there 
seems to be no space to traverse to leave the crystal, it appears to have 
spent some time in the crystal. In fact, the wave packet is brought inside 
the crystal by the pump pulse. This happens when the final state is at a 
band-gap edge, where the local DOS is larger in the crystal than in vacuum, 
see Fig.~\ref{box}(d). Owing to the DOS gradient, the wave packet is shifted 
beneath the surface. Comparison of the streaking by the pulse in Fig~\ref{box}(c) 
and by the pulse of the opposite sign (acceleration at $\tau=0$ vs. deceleration) 
again indicates that the effect is a property of the system rather than of the 
specific perturbation. 

Some features of the $\sigma(\epsilon)$ curves can be readily
understood within the classical picture: the minima B and E and the
maximum C are more pronounced for the deeper defects. The minima 
A and D, however, show the opposite trend. This happens when a rapid 
energy dependence of the photoemission cross-section leads to oscillations 
of $\sigma$. Close to the surface this energy dependence varies with 
the position of the defect, and at certain energies the
effect becomes stronger for the 1st layer than for the deeper layers.
Nevertheless, also in this case the streaking curves retain their 
traditional shape; an example for the topmost layer is shown in 
Figs.~\ref{defect}(c) and~\ref{defect}(d).

These results show that the streaking phase shift may occur at arbitrarily 
small escape depths and without the interference between the layers (in 
the case of the defect it is explicitly excluded). This imposes a limitation 
on using the states located outside the metal as a zero-delay reference: the 
final states must be in an NFE region.

\begin{figure}[t]    
\begin{minipage}{0.45\textwidth}
\includegraphics[width=\textwidth]{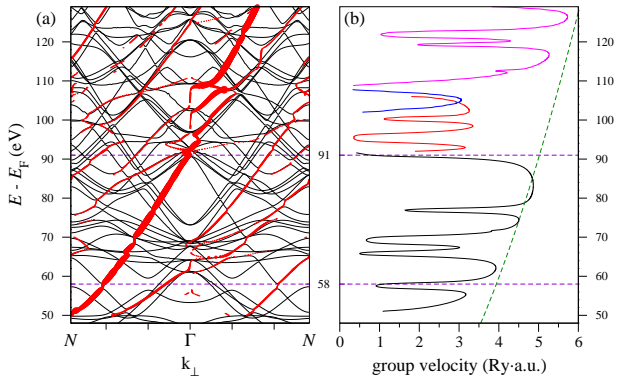}
\end{minipage}
\caption{\label{wolfram} (color online) (a) Real band structure (thin lines)
and conducting complex band structure (red thick lines) of W(110). (b) 
Energy dependence of group velocity for four conducting branches (full lines) 
and in vacuum (dashed line).  
}
\end{figure}

Finally, let us see whether the present theory is relevant to the experiment
on tungsten. The band structure for the normal emission from the (110) surface
($N\Gamma N$ line) is shown in Fig.~\ref{wolfram}. To determine the Bloch waves 
that effect the electron escape into vacuum we draw on the one-step theory of 
photoemission, in which the final states are time reversed LEED states (low 
energy electron diffraction). The LEED wave functions are calculated from the 
complex band structure of the semi-infinite W(110) crystal with the linear 
augmented plane wave method as explained in Ref.~\cite{KS99}. The Bloch waves 
most strongly contributing to the LEED state -- {\it the conducting branches} 
-- are highlighted by the line thickness, which is proportional to the current 
carried by the individual partial wave (see Ref.~\cite{parw} for the explanation). 
The energy dependence of the group velocity for the four conducting branches is 
shown in Fig.~\ref{wolfram}(b). The final states for photoemission from both 
$4f$ and $5d$ bands are seen to fall in the vicinity of the band gaps, at 58 
and 91~eV, respectively. Thus, the effect of advance must occur in both cases, 
and, judging by the variations of $\sigma$ given by the present model (as large 
as 150~as), it strongly contributes to the observed effect. Qualitatively, one 
may speculate that because the velocity mismatch is larger at higher energies 
the negative shift of the streaking curve should be larger for $5d$ than for 
$4f$ emission, which would give an advance of the $5d$ emission, as in the 
experiment. 

To summarize, the final states band structure causes a $\tau$-shift of the 
spectrogram, and its sign depends on whether the photoelectron moves in the 
crystal faster or slower than in vacuum. For an electron coming from a large 
depth and moving faster than in vacuum the streaking curve shifts towards 
positive delays. In the present 1D model (consistent with realistic 
$\mathbf k_\parallel$  projected band structure) the delay is within 50~as. 
This effect is in accordance with classical dynamics. Naturally, it steadily 
decreases with decreasing escape depth. The quantum nature of the outgoing 
electron becomes important for final energies sufficiently close to a band 
gap, where it always moves slower than in vacuum. In that case the spectrogram 
shows an advance. The quantum effect is found to exceed 100~as even for 
electrons excited from the states localized at the surface. This reveals
an aspect important for various electron spectroscopies: the outgoing 
electrons are sensitive to the substrate band structure no matter how 
small the electron mean free path is.

The main implication for the attosecond spectroscopy 
is that neither the inhomogeneity of the streaking field nor an extended 
initial state are a prerequisite for an appreciable relative delay. Still, 
the mean free path plays a large role because depending on its value different 
aspects of the mechanism become important. Experimentally, the predicted 
effect should manifest itself as a non-monotonic dependence of $\Delta\tau$ 
on the XUV energy, and it must be allowed for in interpreting spectrograms. 
At the same time, measurements at different $\omega$ would be instrumental 
in separating the final-states effect from the effect of initial state and 
of the screening of the laser field and providing information on the factors involved.
 
The author gratefully acknowledges fruitful discussions with P.M.~Echenique 
and A.K.~Kazansky. The work was supported in part by the Spanish Ministerio 
de Ciencia e Innovaci\'on (Grant No. FIS2010-19609-C02-02).


\begin{thebibliography}{12}
\bibitem{cmt1} The implications of the finite spectral width of the
               XUV pulse for the streaked spectra are discussed in
               Ref.~\cite{KB07}.
\end{thebibliography}
\end{document}